# Advancing High-Throughput Combinatorial Aging Studies of Hybrid Perovskite Thin-Films *via* Precise Automated Characterization Methods and Machine Learning Assisted Analysis

Alexander Wieczorek[a], Austin G. Kuba[b], Jan Sommerhäuser[a], Luis Nicklaus Caceres[a], Christian Wolff[b,*], Sebastian Siol[a,*]

[a]*Laboratory for Surface Science and Coating Technologies, Empa − Swiss Federal Laboratories for Materials Science and Technology, Switzerland.*

[b]*Institute of Electrical and Microengineering (IEM), Photovoltaic and Thin-Film Electronics Laboratory, EPFL − École Polytechnique Fédérale de Lausanne, Switzerland.*

## Abstract

To optimize materials stability, automated high-throughput workflows are of increasing interest. However, many of those workflows use processes not suitable for large-area depositions which limits the transferability of results. While combinatorial approaches based on vapour-based depositions are inherently scalable, their potential for controlled stability assessments has yet to be exploited.

Based on MAPbI$_3$ thin-films as a prototypical system, we demonstrate a combinatorial inert-gas workflow to study materials degradation based on intrinsic factors only, closely resembling conditions in encapsulated devices. Through a comprehensive set of automated X-Ray fluorescence (XRF), X-Ray diffraction (XRD) and UV–Vis characterizations, we aim to obtain a holistic understanding of thin-film properties of pristine and aged thin-films. From phase changes derived from XRD characterizations before and after aging, we observe similar aging behaviours for MAPbI$_3$ thin-films with varying PbI$_2$ residuals. Using a custom-designed *in-situ* UV–Vis aging setup, the combinatorial libraries are exposed to relevant aging conditions, such as heat or light-bias exposure. Simultaneously, UV-Vis photospectroscopy is performed to gain kinetic insights into the aging process which can be linked to intrinsic degradation processes such as autocatalytic decomposition.

Despite scattering effects, which complicate the conventional interpretation of *in-situ* UV–Vis results, we demonstrate how a machine learning model trained on the comprehensive characterization data before and after the aging process can link optical changes to phase changes during aging. Consequently, this approach does not only enable semi-quantitative comparisons of materials' stability but also provides detailed insights into the underlying degradation processes which are otherwise mostly reported for investigations on single samples.

## 1. Introduction

The advancement of energy conversion and sensory technologies relies heavily on developing novel semiconducting materials, with innovative functional properties. A class of hybrid organic-inorganic semiconductors that garnered attention for high-performance applications in light-emission,[1] photovoltaics,[2] and X-Ray detection[3] are metal halide perovskites (MHP). MHPs crystallize in the perovskite structure and follow the general formula ABX$_3$, where typically A = Cs$^+$, methylammonium (MA$^+$), formamidinium (FA$^+$), B = Sn$^{2+}$, Pb$^{2+}$, and X = I$^-$, Br$^-$, Cl$^-$.

A key challenge in the development of hybrid semiconductors is their limited long-term stability under operating conditions.[4,5] Although multiple concepts exist to predict the phase stability of hybrid perovskites, *e.g.* through the long-established Goldschmidt tolerance factor,[6] and octahedral tilting due to the A-site cation,[7–10] decomposition may occur *via* a multitude of pathways. This includes thermal instability,[11] photoinduced degradation,[12] as well as oxygen- and moisture-induced degradation.[13] Consequently, experimental assessments are necessary to understand their degradation behaviour and unlock their true potential for the aforementioned applications on the commercial scale.

Nonetheless, performing aging studies can be especially challenging for multiple reasons:

1) Aging is intrinsically time-intensive, even under accelerated application-specific aging conditions.[14,15]
2) Assessing aging using only one material characterization method typically yields limited insights into the complex changes in the material chemistry which may result in wrong interpretations.
3) Aging may result from multiple intrinsic and extrinsic factors at once, rendering the assignment to one specific factor difficult. At the same time, different extrinsic factors can limit the reproducability across laboratories.
4) The durability of a material can be highly dependent on the deposition technique.[16] Consequently, the results from stability screening may differ significantly for different synthesis routes.

Mainly the first issue has been increasingly targeted through various approaches. This includes parallelized high-throughput experiments (HTE) on photovoltaic cells under accelerated aging conditions[17] as well as automated materials acceleration platforms for screening the optical stability of thin-films.[18] While these strategies allow one to significantly reduce the time effort, there is a critical demand to address the remaining challenges.

A more established approach for experimental thin-film screening using scalable deposition methods are vapour-based combinatorial techniques.[19] Using such techniques, material libraries can be prepared, which exhibit process parameter gradients or composition gradients across a single substrate. This way large areas of the synthesis phase space are accessed in one deposition run. Driven by advancements in experimental infrastructure and data processing,[20] automated characterization techniques have greatly improved in recent years,[21] yielding more thorough insights into the respective materials properties across libraries. Although first reports have been made to use combinatorial approaches for the optimization of devices,[22] its potential for stability screening with respect to compositional stability has yet to be exploited.

In this work, we report a comprehensive workflow for combinatorial thin-film aging studies under highly controlled environmental conditions. As extrinsic factors are less relevant for the commercialization of MHPs due to modern encapsulation techniques,[23,24] we exclude extrinsic aging factors in our screening protocol. This is enabled through a comprehensive inert-gas workflow, in which the samples remain in nitrogen atmosphere during XRF, XRD and UV–Vis characterization, as well as the transfer between instruments.

We demonstrate our workflow based on $MAPbI_3$ as a prototypical system. Since its degradation mechanisms have already been investigated in detail, it serves as an ideal reference material. Furthermore, we employ a scalable deposition technique based on a two-step process that holds promise for the conformal deposition of these semiconductors on large areas. Using a novel custom-designed platform for combinatorial *in-situ* UV–Vis aging experiments, we investigate how conversion gradients introduced during this process affect the resulting optical stability of the thin-films. While initial analysis of the UV-Vis data cannot be directly linked to the aging trends revealed from XRD measurements of the pristine and aged thin-films, machine-learning assisted analysis of the UV-Vis spectra allows us to reveal common aging kinetics for the herein investigated thin-films.

Overall, the results of this study highlight how a comprehensive stability-screening workflow performed under inert-gas atmosphere can generate valuable insights into the materials aging behaviour under application-relevant degradation conditions. While our investigation is focused on MHPs, the framework may also be used to accelerate the development of other promising material systems as stability assessments are crucial for a wide range of novel materials.

# 2.    Experimental section

## Materials

Lead iodide (PbI$_2$, 99.999%) beads were purchased from Sigma Aldrich. Methylammonium iodide (MAI, >98%) was purchased from Greatcell Solar Ltd. Isopropanol ($^i$PrOH, 99.5%, anhydrous) was purchased from Sigma-Aldrich. All materials were used without further purification.

## Sample preparation

The samples were deposited using a two-step technique which consists of the deposition of an inorganic BX$_2$ template (*here:* PbI$_2$) followed by coating of the organic AX salt (*here:* MAI) as further explained below.

### *Inorganic template preparation*

The inorganic templates were prepared via thermal evaporation in a Lesker Mini SPECTROS evaporation system. 1.1 mm thick borosilicate glass (EXG) substrates of 50 · 50 mm$^2$ size were ultrasonically cleaned in acetone, and ethanol, followed by drying in a flow of N$_2$. The substrates were then transferred into the evaporation system, substrate masks were used to generate a patterned template. With closed shutters, the PbI$_2$ source was first heated until a stabilized rate of 1 Å s$^{-1}$ was acquired. Then, the shutters were opened and PbI$_2$ was deposited at 1 Å s$^{-1}$ until a nominal film thickness of 200 nm was determined by a quartz crystal monitor (QCM). The substrate was intentionally not rotated during this process to induce thickness gradients resulting in conversion gradients from the PbI$_2$ to the resulting perovskite.

### *Deposition of organic precursors*

Previously prepared inorganic templates were transferred to another glovebox without breaking inert-gas conditions and all further steps were performed within this glovebox. The MAI solution was first prepared by dissolving 300 mg within 5 ml $^i$PrOH. The mixture was first vortexed for 1 h and then filtrated using a 45 µm polytetrafluoroethylene (PTFE) syringe filter. All used solutions and solvents were then pre-heated at 70 °C for 20 min. As a pre-wetting step, 1 ml preheated $^i$PrOH was first deposited onto the inorganic templates and left resting for 1 min and finally spun at 1300 rpm for 10 s. 1 ml of the preheated MAI solution was then deposited and left resting on the template for 1 min. To promote the solution evaporation, the sample was then spun at 1300 rpm for 30 s. Films were then annealed at 100 °C for 10 min. Residual organic salts were then removed through the dynamic addition of 500 µl $^i$PrOH at 2500 rpm. To reduce the amount of surface defects on the MAPbI$_3$ surface without affecting the crystal lattice, a MAI-based surface treatment was applied as reported elsewhere.[25]

## Characterization techniques

### *X-Ray diffractometry (XRD)*

XRD measurements were performed in a Bruker D8 Discovery diffractometer using Cu kα radiation in Bragg-Brentano geometry. Within a glovebox, the sample was introduced in a homemade gas-tight XRD sample chamber equipped with a 1 mm thick Polyether ether ketone (PEEK) dome. The dome material was chosen due to its high X-Ray transmission properties compared to other widely available polymers.[26] For each sample spectra, a reference spectra recorded using an empty dome was subtracted. Batch analysis of the XRD peaks for fitting was further performed using the open source package COMBIgor[27] within an Igor Pro 9 environment. 2$\Theta$ scans were performed within the 5 – 50° range. $X$ scans were performed by first setting the 2$\Theta$ values to the maximum positions of the PbI$_2$ (001) and MAPbI$_3$ (110) features at 12.65° and 14.1°, respectively. The $X$ angles were then scanned in the -5 – 90° range during the measurement.

For the semi-quantitative analysis, reference crystal structures were first retrieved from the Inorganic Crystal Structure Database (ICSD).[28] Using ICSD-68819(ICSD release 2023.2) for $PbI_2$ and ICSD-124919(ICSD release 2023.2) for $MAPbI_3$, structure factors $|F|$ were extracted with VESTA 3.5.8 and divided by their respective number of formula units $z$. The resulting values were $|F_{PbI_2(001)}|$ = 63, and $|F_{MAPbI_3(110)}|$ = 102. Furthermore, the multiplicities $p$ of the reflection planes were $p_{PbI_2(001)}$ = 2, and $p_{MAPbI_3(110)}$ = 4.

To account for intensity variations resulting from different orientations of the sample specimen relative to the incident beam, the Lorentz factor $L$ is further considered in the semi-quantitative analysis of the XRD data. It is defined as:[29]

$$L = \frac{1 + \cos^2(2\theta)}{\sin^2(\theta) \cdot \cos(\theta)} \quad (1)$$

, where $\theta$ denotes the Bragg angle of the observed reflection.

## *X-Ray fluorescence (XRF) spectroscopy*

XRF measurements were performed in a Fischerscope XDV-SDD X-ray fluorescence (XRF) system equipped with an Rh X-ray source to determine the atomic ratios of Cs, Pb, I and Br. The thickness of each thin-film was further estimated from the attenuation of signals related to the sample substrate. To maintain all samples under inert-gas conditions, 3D printed containers covered with a thin polyethylene (PE) foil were used. Measuring through this PE foil notably did not significantly affect the measured thickness due to the low sensitivity of XRF for 2nd row elements and minimal attenuation of the underlying signal.

## *In situ UV–Vis spectroscopy*

*In-situ* UV–Vis measurements were performed in a homemade setup. The combinatorial library, contained within the climate chamber, was mounted on a biaxial stepper motor stage (Thorlabs LTS300). During measurements, the sample was illuminated using an Deuterium-Halogen light source (Ocean Optics DH-2000-BAL). Two charge-coupled device (CCD) detectors (Ocean Optics HR4Pro XR-ES) were used to record the resulting transmission and reflection spectra in the 200 nm to 1100 nm range. A 50:50 polka dot beam splitter (Thorlabs BPD5254-G01) mounted at an 45°-angle was used to separate the reflected light from the incoming light beam. Solarization-resistant multi-mode fibres (Thorlabs M112L02) with a core diameter of 200 μm were used to route the optical beams. All fibres were terminated with FOV-adjustable collimator lenses (Ocean Insight 74-ACR), mounted on biaxial kinematic mounts for aligning the beams. Two plano-convex focusing lenses (Thorlabs LA4052) were used to collimate light being reflected and transmitted through the sample. All components mentioned above are specified for use within the optical range from 200 nm to 1100 nm or better, allowing measurements across the entire UV–VIS spectrum. The beam path for the reflection measurement first passes straight through the beam splitter and the focusing lens, is reflected at the sample back through the lens and is then reflected by the beam splitter at an 90°-angle before entering the reflection spectrometer. The beam path for the transmission measurement passes straight through the beam splitter, the focusing lens, the sample and the second focusing lens before entering the transmission spectrometer on the other side of the sample (Figure S 1). The entire measurement setup is enclosed in a box made from black anodized aluminium to block stray light.

Based on the obtained reflectance and transmission spectra, the reflection-corrected transmission $T'$ was calculated according to:

$$T' = \frac{T}{1 - R} \quad (2)$$

, where $T$ denotes the sample transmission and $R$ denotes the measured sample reflection (Figure S 2). Based on this value, the absorption coefficient $\alpha$ of the thin-film layer was calculated:

$$\alpha = \frac{-\ln\left(\frac{T'_{\text{Substrate+Film}}}{T'_{\text{Substrate}}}\right)}{d} \quad (3)$$

, where *d* denotes the thin-film thickness estimated from XRF measurements.

To ensure inert-gas conditions are maintained during the measurement, a homemade climate chamber was used. It was continuously held at an overpressure of approximately 20 mBar using a continuous $N_2$ inflow and a one-way valve with a corresponding cracking pressure. To enable aging under stress-light conditions, a 70 · 70 $mm^2$ white-light LED chip was used in combination with a light diffuser. Based on the characterized emission spectrum, the light intensity was calibrated to 1 kW $m^{-2}$ using a Si photodiode with a known external quantum yield (EQY) (Figure S 3). The spatial uniformity of our white light-stressing conditions were characterized to be better than ±1.4% across the area of our combinatorial libraries in combination with a high temporal stability (Figure S 4). The entire climate chamber body is heated to 85 °C using two heating cartridges connected to a PID Controller (Red Lion PXU11A50). From infrared camera measurements of the thin-films, a heating uniformity of ±1 K was confirmed (Figure S 5). As a result of these highly accurate aging conditions across the library area, reliable combinatorial aging is enabled. Overall, the aging conditions reflect the ISOS-LC3I standard.[15] Homemade code implemented in LabView 2021 was used to automate the measurements.

## Machine learning analysis

The machine learning model was implemented using the open-source python packages XGBoost[30] and Scikit-Learn[31] within a Python 3.7.1 environment. The data set was first shuffled and split it into training and testing data sets with a ratio of 8:2. Then, hyperparameter optimization was performed using 3-fold cross validation using root-mean-squared-error (RMSE) as the scoring metric. To prevent overfitting on our experimental data set, we limited the optimizable hyperparameter space to a low maximum tree-depth of 2 – 4, a low learning rate of $10^{-3} – 10^{-1}$, and a high amount of estimators of 200 – 1000. Furthermore, the regularization parameters Gamma, Alpha and Lambda have been optimized. A gamma value of 0.7 was found to be optimal, in agreement with the low maximum tree-depth. Lastly, the performance of the model was assessed through prediction on the earlier split test set using $R^2$ and *RMSE* as metrics.

## 3. Results and Discussion

### Combinatorial workflow

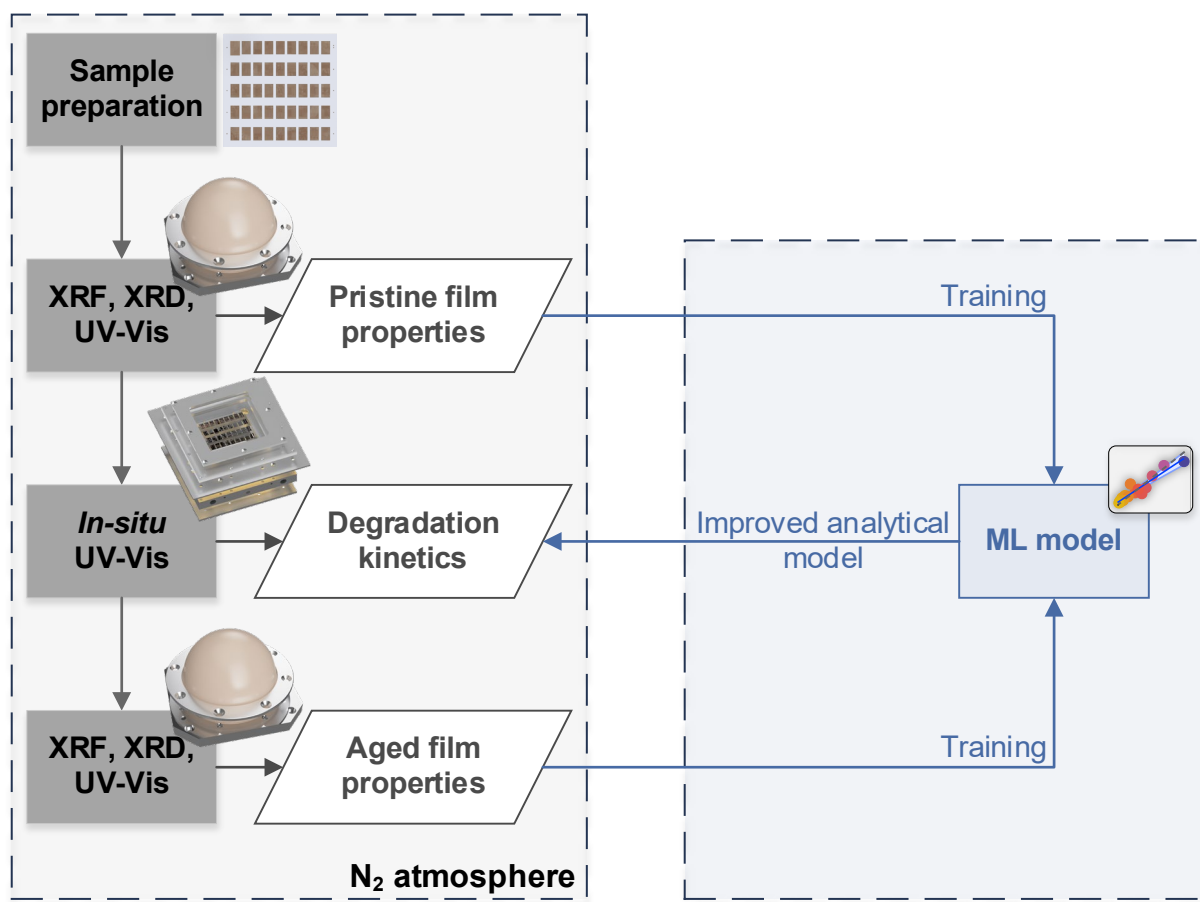

**Fig. 1. Inert-gas synthesis & characterization workflow for combinatorial perovskite libraries.** The rendered graphics within the inert-gas workflow depict a combinatorial perovskite library, the inert-gas XRD Dome, the climate chamber for UV-Vis aging, and repeatedly the inert-gas XRD Dome (top to bottom). The combined data sets obtained from the characterization of the pristine and aged thin-films are then used to train a machine learning (ML) model for the analysis of the optical spectra obtained during the *in-situ* UV-Vis aging characterization.

Central targets of our combinatorial workflow have been the integration of an existing scalable deposition processes and the exclusion of external degradation effects during aging and characterization. To address the first aspect, we employ gradient deposition as the combinatorial deposition technique which has previously been demonstrated for scalable vapour- and solution-based deposition techniques.[21] Specifically, we employ a two-step deposition process consisting of thermal evaporation of an inorganic $PbI_2$ template, followed by solution processing and annealing of the organic MAI salts. This approach was previously shown to be ideal for conformal coatings on micro-structured substrates.[32] Combined with its high potential for upscaling using slot-die coating and mitigation of toxic solvents such as DMF,[33,34] it is highly promising for an industrial scale deposition process, especially for other MHPs suitable for integration with Si-based optoelectronics (*e.g.* perovskite-Si tandem solar cells).[35] During this process, underconversion (*i.e.* incomplete formation of the MHP from the inorganic $BX_2$ template) or overconversion (*i.e.* the formation of an AX-rich phase) may easily occur.[36,37] To investigate the effect of these conversion gradients on the resulting aging behaviour, we deliberately introduce gradient thicknesses on the inorganic template. Furthermore, to improve the validity of results, we employ patterned libraries during synthesis, thereby preventing lateral aging effects across different samples (*e.g. via* lateral diffusion or local $PbI_2$ outgassing).[38,39]

To exclude external degradation the entire characterization process is performed under an inert-gas atmosphere. This is crucial to screen for intrinsic MHP stability,[15] which is important for implementation in encapsulated devices, and is typically not considered for stability screening workflows. To this end, we developed custom solutions to perform X-Ray and optical-based methods under inert-gas: First, a comprehensive thin-film characterization is performed based on XRD, XRF and UV-Vis measurements, allowing to deduce structural information, thin-film thickness and the optical properties of the pristine thin-films. Then, *in-situ* UV-Vis measurements are performed under highly-controlled aging conditions (85 °C, 1 kW m$^{-2}$ white light, N$_2$ gas) to promote the intrinsic aging of the thin-films while rapidly probing changes in the optical properties. Lastly, the comprehensive sample characterization initially performed on the pristine library is repeated on the aged library.

Based on these comprehensive experimental data sets of pristine and aged thin-films, we are able to leverage data-driven approaches to enable a better understanding of the observed optical changes with respect to phase changes. This knowledge can then be used to provide a better analytical model for the *in-situ* UV-Vis spectra and understand the underlying aging kinetics.

## Effect of conversion ratio on pristine thin-film properties

Based on the characterization of the pristine thin-films, the effect of varying PbI$_2$ residuals across the library on the optical properties can be determined.

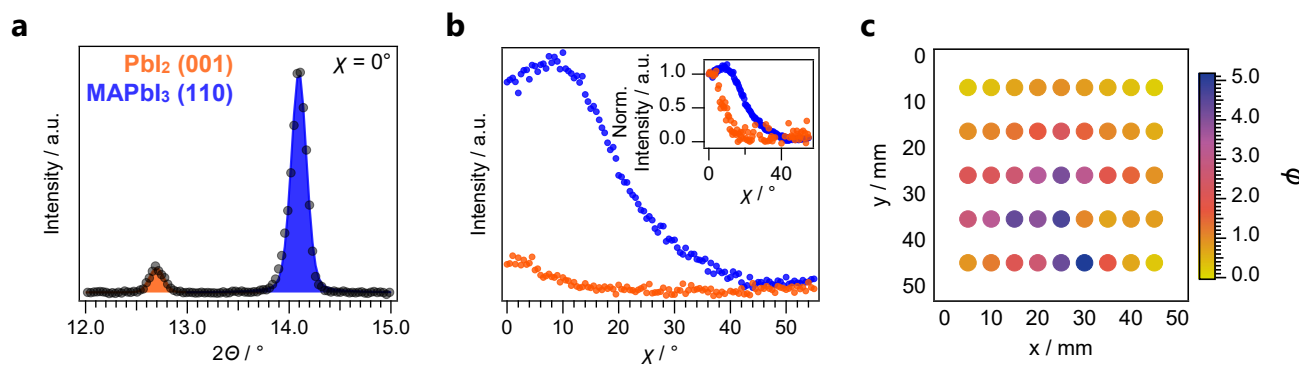

**Fig. 2. XRD characterization of the combinatorial MAPbI$_3$ sample**. a) 2-Theta scan for a selected sample and b) Azimuthal $X$-scan for the same sample. The inset depicts traces that have been normalized at $X = 0°$, according to the term defined in Eq. 3. Based on these individual results, the conversion ratio is calculated for each sample. c) Conversion ratio $\varphi$ as function of the position on the combinatorial library.

First, the amount of residual PbI$_2$ within the MHP thin-film was approximated using XRD measurements. This is a common approach where typically the most prominent peak intensities of PbI$_2$ and the corresponding MHP compared (Fig. 2.a).[40–43] We further confirmed the absence of over converted MHP phases through the lack of features at shallower $2\Theta$ angles (Figure S 6).

Thin films produced by PVD often exhibit pronounced texture. Especially in oblique-angle vapour-deposition setups, grains tend to grow towards the source.[44,45] Since the commonly employed Bragg-Brentano XRD measurements only probe features that align with the diffraction plane, preferential orientation can result in intensity variations, which are not correlated with the phase constitution of the material. To account for these effects, we performed azimuthal scans on PbI$_2$(001) and MAPbI$_3$(110) features (Fig. 2.b) to probe the preferred orientation of each phase and account for geometry-specific intensity variations.[46]. Compared to PbI$_2$ which exhibited a clear out-of-plane orientation, the resulting MAPbI$_3$ thin-films were less textured. These findings are in agreement with previous results from synchroton-based Grazing-Incidence Wide-Angle X-ray Scattering (GIWAXS) measurements on two-step deposited MHPs grown from thermally evaporated templates.[47,48] As a result of the the difference in texture between the two phases, determination of the phase fraction from the $2\Theta$ scans alone would result in an underestimation of the MHP phase ratio compared to PbI$_2$. To account for this effect, the normalized peak area $A_{\text{norm}}$ is first calculated according to:

$$A_{\text{norm}} = A_{2\Theta} \cdot \frac{A_X}{I_{X=0°}} \qquad (4)$$

, where $A_{2\Theta}$ denotes the peak area obtained from the $2\Theta$ measurement, $A_X$ denotes the area under curve for the $X$ scan performed at the given reflection and $I_{X=0°}$ denotes the absolute intensity at $X = 0°$. The latter factor effectively resembles the previously depicted normalized $X$ scans (inset in Fig. 2.b).

For measurements performed at room temperature, the fraction of chemical formal units of MAPbI$_3$ against PbI$_2$ may be approximated as the conversion ratio $\varphi$:[29]

$$\varphi = \frac{\frac{A_{\text{norm, MAPbI}_3(110)}}{|F_{\text{MAPbI}_3(110)}|^2 \cdot p_{\text{MAPbI}_3(110)} \cdot L_{\text{MAPbI}_3(110)}}}{\frac{A_{\text{norm, PbI}_2(001)}}{|F_{\text{PbI}_2(001)}|^2 \cdot p_{\text{PbI}_2(001)} \cdot L_{\text{PbI}_2(001)}}} = \frac{A_{\text{norm, MAPbI}_3(110)}}{A_{\text{norm, PbI}_2(001)}} \cdot \frac{|F_{\text{PbI}_2(001)}|^2 \cdot p_{\text{PbI}_2(001)} \cdot L_{\text{PbI}_2(001)}}{|F_{\text{MAPbI}_3(110)}|^2 \cdot p_{\text{MAPbI}_3(110)} \cdot L_{\text{MAPbI}_3(110)}} \quad (5)$$

, where $F$ further denotes the structure factor per formal unit for the given reflection plane, $p$ denotes the multiplicity of the given reflection plane within each crystal structure, and $L$ defines the Lorentz-polarization factor, a trigonometric term considering intensity changes resulting from the orientation of the sample towards the incident beam (see method section). Errors in this approximation may arise from amorphous phases within the thin-film or the significant presence of a layered structure within the thin-film. However, as these factors would be expected to be similar for all samples, they result in a systematic error of $\varphi$. Consequently, it serves as a suitable semiquantitative descriptor for the phase ratio within the studied thin-films.

The investigated combinatorial library exhibits a large range of conversion ratios ranging from $\varphi \approx 0.2$ to $\varphi \approx 5$, (Fig. 2.c), rendering it ideal for the combinatorial analysis of PbI$_2$ residuals within MAPbI$_3$. From repeated measurements of the same sample after extended periods of time within the inert-gas XRD solution, we further verified the phase stability of the library during this characterization (Figure S 7).

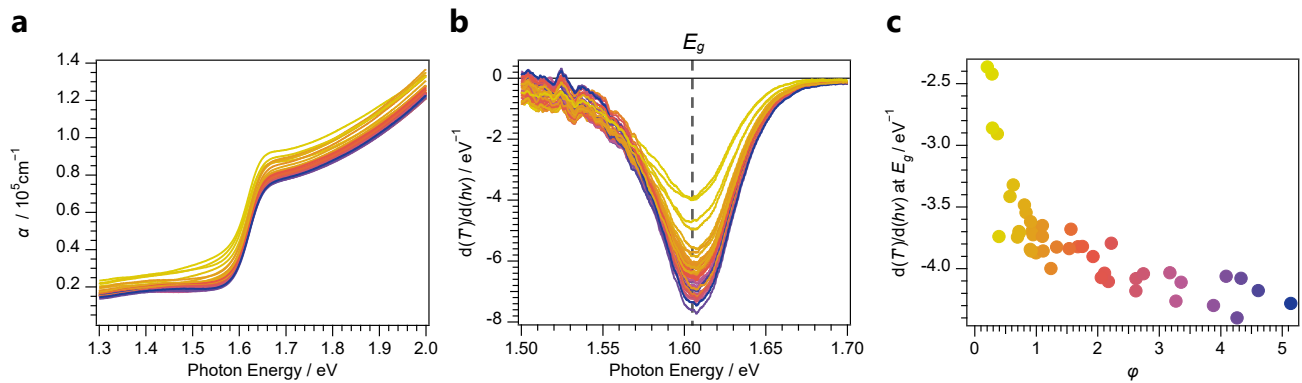

**Fig. 3.** a) Absorption coefficients $\alpha$ with varying $\varphi$. b) First derivative of the reflection-corrected transmission $T'$, showing the steepness of the transmission onset near the band gap. c) Extracted values of subplot b, indicating a clear correlation between the MHP/PbI$_2$ and the steepness of the transmission onset near the bandgap.

Although not clearly visible by eye (Figure S 8), subtle changes in optical properties across the library could be successfully resolved from automated UV-Vis measurements. A clear trend with respect to the phase ratio was observed: Samples with higher MHP phase contents exhibited an overall reduced absorption below and above the bandgap of 1.6 eV (Fig. 3.a) compared to those with lower MHP phase fractions. The PbI$_2$ template deposited by thermal evaporation did not exhibit notable optical absorption within this range (Figure S 9), indicating that optical scattering from mixed phases may cause this effect. Notably, the intensity of the first derivative of the reflection-correct sample transmission $T'$ (Fig. 3.b), and hence the steepness of the optical absorption, is highly correlated to $\varphi$. In contrast, the maximum exhibited shifts of less than 0.1 eV and is therefore less sensitive to the changes of $\varphi$. Likewise, manually extracted band gaps by Tauc plot analysis varied by less than 0.2 eV across the complete data set (Figure S 10), which lies below inaccuracies typically resulting from automatic Tauc-plot analysis algorithms.[49] Hence, observation of the changed steepness of the absorption onset appears more robust than extracted band gaps to track phase-composition changes within the thin-film. Furthermore, these results highlight how a comprehensive automated thin-film analysis allows to resolve trends otherwise not obtainable from either experiments on few samples or HTE workflows relying on optical measurements alone.

## Effect of conversion ratio on aging

To experimentally screen hybrid perovskite thin-films with respect to their phase stability, RGB-camera-based optical characterizations have previously been successfully employed.[50] Despite first successful demonstrations of image-based spectra-prediction for semiconductors by Stein *et. al*,[51] reliably resolving finer spectral changes within the optical absorption from this approach remains elusive.

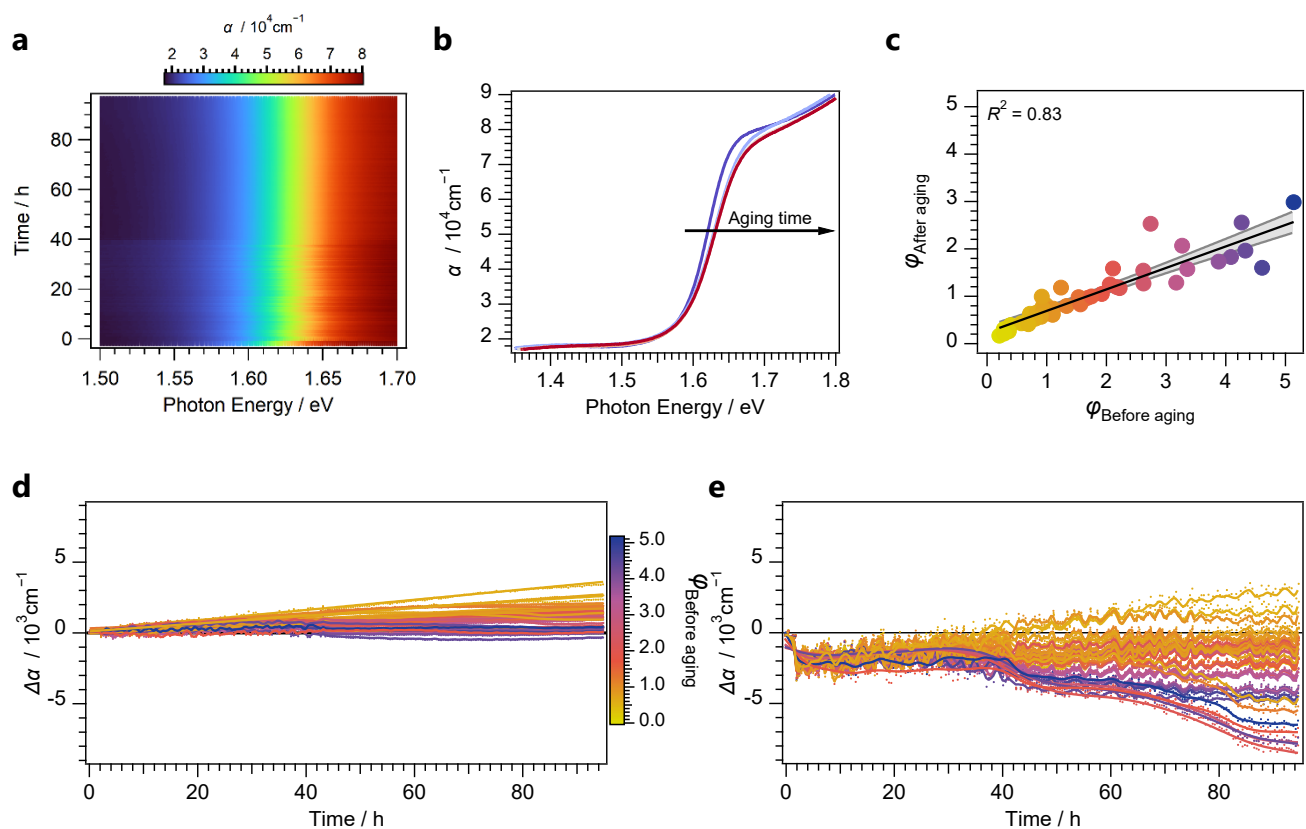

**Fig. 4.** a) Change of optical properties over time for a single sample. b) Selected spectra for a single sample at equidistantly spaced aging time intervals. c) Results from phase ratio determination before and after aging process, indicating a linear loss across the library. d) Changes in absorption below the bandgap (at 1.4 eV). e) Changes in optical absorption above the bandgap (at 1.7 eV).

Indeed, despite the volatile nature of the MA$^+$ cation, we observed minute spectral changes during aging for thin-films with higher $\varphi$ under inert conditions (Fig. 4.a). Similar to the differences between pristine thin-films with varied $\varphi$, we observed a reduction in the absorption onset slope and an increase in the sub-bandgap absorption (Fig. 4.b). These trends are aligned with *in-situ* UV-Vis aging studies performed for single MHP samples.[52]

The back conversion from the MHP to PbI$_2$ was further confirmed through post-mortem XRD measurements. Remarkably, we determined the change in conversion to be equal for all samples as evident from the linear relationship between MHP/PbI$_2$ Phase ratios $\varphi$ before and after the aging process (Fig. 4.b). Through observation of absorption coefficient changes $\Delta\alpha$ both below (at 1.4 eV, Fig. 4.d) and above the bandgap (at 1.7 eV, Fig. 4.e), the evolution of optical properties as a function of aging time can be visualized across the library.

Below the bandgap, a linear increase in the absorption coefficient of up to $\Delta\alpha = 3 \cdot 10^3$ cm$^{-1}$ could be observed over the aging time. This increase was particularly profound for samples exhibiting low MHP/PbI$_2$ phase ratios, indicating an increase in optical light scattering. Scattering effects in thin-films can have multiple origins, such as interfacial roughness, as well as interference effects, and bulk scattering.[53] Thus, a quantitative understanding would require further optical modelling which is beyond the scope of the present work.

Still, these trends enable a qualitative understanding of the complex changes occurring above the bandgap: Initially, a decrease in the optical absorption within 10 h of aging was observed for all samples. Afterwards, while low converted samples exhibited a clear increase in absorption from likewise increased scattering, high converted samples exhibited a decrease in the absorption coefficient, resulting in a range of $\Delta\alpha = 11 \cdot 10^3$ cm$^{-1}$. These effects can likely be assigned to an overlay of scattering effects as well as a loss of optical absorption from outgassing of the organohalides. Interestingly, the observed trends follow a less linear trend with respect to the aging time, indicating a non-linear trend for the outgassing. These kinetics are most pronounced for the highly converted samples where the increase of the optical scattering during the aging time was observed to be negligible below the bandgap. The s-shaped feature in the optical absorption changes above the band gap at 40 h can be linked to autocatalytic degradation resulting from the local formation of $I_2$ from HI vapour during aging.[54] This resembles recent reports where similar trends were determined from XRD measurements with lower temporal resolution.[42]

Overall, these findings showcase the complex and nonlinear changes in optical spectra that occur during the degradation process of MHP thin-films and result in implications for their optical stability screening: The apparent optical absorption of MHP thin-films above the band-gap during aging can be heavily influenced by scattering effects, complicating the quantitative analysis. While previous approaches using RGB-cameras were able to deduce similar kinetic results from red color channels alone,[55] these effects are critical for samples with overall lower optical absorption than the herein investigated thin-films with $PbI_2$ residuals. Furthermore, the autocatalytic nature of the herein observed degradation mechanism induces nonlinear changes of the optical properties during aging. As a result, linear extrapolation of trends observed only before and after aging may give a falsified indication on the actual materials stability.

## Machine learning assisted analysis of aging kinetics

To gain a more detailed understanding of the degradation kinetics across the complete library, we continued our investigation with data-driven approaches to extract kinetic information from the recorded *in-situ* UV–Vis spectra. The high correlation of trends observed from XRD and UV–Vis measurements motivated us to develop a model that can predict $\varphi$ from UV–Vis spectra obtained during the aging process, effectively yielding insights about the phase composition evolution from our *in-situ* UV–Vis investigation.

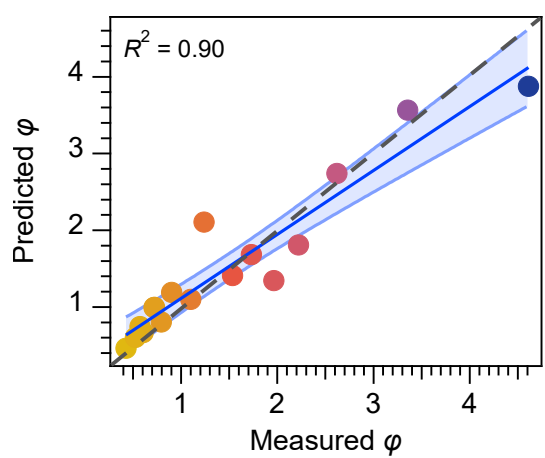

**Fig. 5.** Prediction of the MHP/$PbI_2$ phase ratio $\varphi$ from UV–Vis spectra by the ensemble regression model, achieving a high R$^2$ score of 0.90 and a RMSE of 0.36.

To ensure generalization while still being able to account for non-linear trends occurring in optical absorption spectra, we focused on ensemble methods, striking a balance between simple linear regression and deep learning approaches. Based on a gradient-boosting regressor, optimized for smaller data sets (see method section),[30] we obtained a model which generalized well on the testing data set (Fig. 5), yielding a root-mean-square-error (*RMSE*) of 0.36 for the prediction of $\varphi$.

The feature selection was performed based on the previous empirical trends observed for MHP thin-films with varied $\varphi$: To learn from the complete spectral shape, we first chose the ratio of optical absorption below and above the bandgap as our input feature. Then, in order to better learn from the complete spectral shape, absorption coefficient ratios at 1.7 eV vs 1.5 eV as well as at 2.0 eV vs. 1.4 eV were used. Additionally, the initially introduced maximum from the first derivative of the optical transmission around the MHP bandgap (Fig. 3.b) was included. Analysis of the feature importance revealed the latter feature to be the most relevant, while the two absorption coefficient ratios used to learn from the spectral shape were similarly important (Figure S 11).

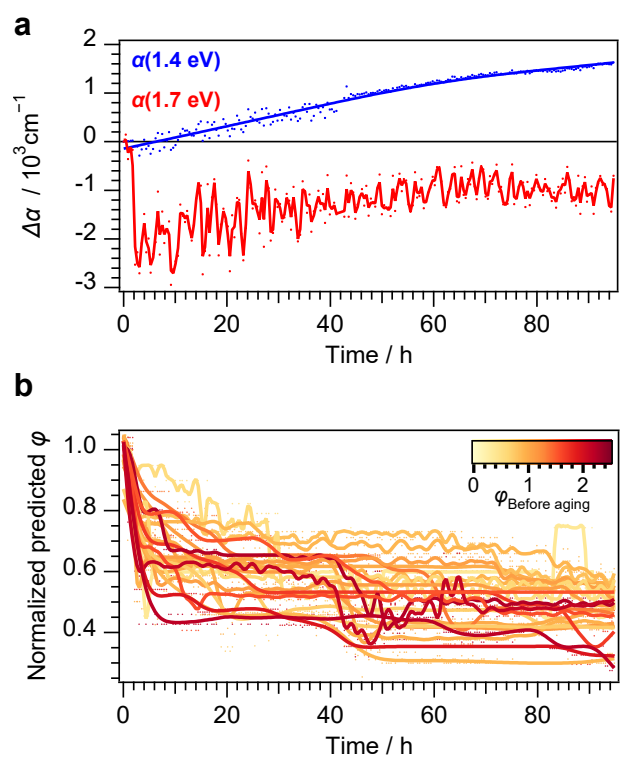

**Fig. 6.** a) Observation of the change in optical absorption $\Delta\alpha$ for a representative sample with $\varphi_{Before\ aging}$ = 0.90 and thus large amounts of scattering effects as evident from the absorption increase below the bandgap. b) Normalized predicted kinetic traces expressed as changes in $\varphi$ for samples with low MAPbI$_3$ phase ratios ($\varphi$ < 2.5), showing a continuous decrease.

When applying this method on our *in-situ* UV–Vis data set, we were able to unravel the kinetics of less converted samples, previously inaccessibe through more classic observations of the optical spectra alone (Fig. 6.b):

A continuous decrease of $\varphi$ and therefore the loss of the MAPbI$_3$ phase during aging could be observed, resulting in a rapid loss of at least 20% from the initial value after 20 h for all studied samples. This is in stark contrast to the smaller decrease in the optical absorption observed for all investigated thin-films above the bandgap and lack of any changes below the bandgap (Fig. 4.e).

Even more strikingly, the s-shaped feature at approx. 40 h of aging time could be observed for most samples. This is in agreement with the autocatalytic degradation at this point and the qualitative observations made from thin-films without significant scattering effects. Fluctuations overlaying these trends could be seen for extremely low-converted samples presented in yellow hues. Most likely, this is a result of the prediction error as samples exhibiting low $\varphi$ values as pristine samples had lower absolute changes of $\varphi$ after aging (Fig. 4.c). Beyond the inflection point of this degradation effect, further decreases stayed within 10% of the initial $\varphi$ value, even when accounting for fluctuations within the predictions. This again highlights the nonlinear nature of the degradation behavior detected from *in-situ* UV–Vis characterization with high temporal resolution.

Interestingly, applying the regressor model on the higher converted samples gave less steady trends (Figure S 12). This is both apparent from higher fluctuations and predictions beyond $\varphi \geq 1$ at 50 h of aging time. These shortcomings can be explained by the training bias as entries with $\varphi > 2.5$ were only found in unaged samples and therefore comprised only 20% of the training data (Figure S 13). Furthermore, despite their high potential for data-driven experimental research,[56–58] tree-based regression models are typically not well-suited for extrapolation.[59]

For the herein employed workflow however, these shortcomings are not crucial since applying the machine-learning analysis is of most use for degraded and low converted samples that are typically more prevalent within the training data. Thus, our findings highlight both the utility and limitations of applying these models for the quantitative analysis of UV–Vis data for materials' degradation analysis.

## Conclusions & Outlook

A comprehensive automated inert-gas workflow was developed and employed to investigate the aging of MHP thin-films with various amounts of $PbI_2$ residuals using a combination of XRF, XRD and UV–Vis characterization methods. From combinatorial *in-situ* UV–Vis aging studies, we obtained a detailed understanding of the underlying degradation kinetics. While the overall mechanism was in alignment with an autocatalytic degradation mechanism as reported elsewhere for experiments not performed under-inert gas,[42,55] we were able to detect the independence of the degradation kinetics from the phase composition for underconverted MHP samples. We further explored how machine-learning assisted analysis of UV–Vis data with featurization based on a detailed materials characterization before and after the aging experiments can be used to link results from optical analysis to conversion rate. This enables deeper insights into the degradation kinetics beyond qualitative discussions alone. The plateauing effect observed for all investigated thin-films within this study suggests a common self-limiting mechanism for their aging beyond the inflection point which could be of interest for further investigations.

Overall, beyond the aforementioned increased understanding of the MHP thin-films aging kinetics, our results highlight how optical screenings of thin-films can be further improved: Relying on optical changes alone, especially at one wavelength only, may provide limited insights of the actual phase stability within the thin-films. These effects would be especially pronounced for material systems exhibiting lower optical absorption coefficient maxima, *e.g.* charge-transport-layers, transparent conductive films, or antireflective coatings, but can be minimized by analysis of the complete spectral shape. For selected material systems, the simulatenous measurement at few selected wavelengths, *e.g.* below, around and above the bandgap, may be sufficient. Including complementary characterization techniques within the workflow can provide crucial validity checks and enable a more detailed scientific understanding of the underlying aging mechanisms. The usage of machine-learning assisted analysis techniques can aid the experimental researcher to extract actionable intelligence from the large amounts of data generated by these workflows. Lastly, the herein presented workflow is ideal for integrating other techniques such as X-Ray photoelectron spectroscopy (XPS) for surface analysis, which would provide a deeper understanding of materials aging inaccessible from few experiments alone.[60,61]

## Author Contributions

Alexander Wieczorek: Conceptualization, Formal Analysis, Investigation, Methodology, Software, Validation, Visualization, Writing – Original Draft, Writing – Review and Editing. Austin G. Kuba: Investigation, Methodology, Supervision, Writing – Review and Editing. Jan Sommerhäuser: Software, Writing – Review and Editing. Luis Nicklaus Caceres: Software, Writing – Review and Editing. Christian Wolff: Funding Acquisition, Resources, Supervision, Writing – Review and Editing. Sebastian Siol: Conceptualization, Investigation, Methodology, Funding Acquisition, Project Administration, Resources, Supervision, Writing – Review and Editing.

## Conflict of Interest

The authors declare no conflict of interest.

# Acknowledgements

All authors acknowledge funding from the Strategic Focus Area–Advanced Manufacturing (SFA–AM) through the project Advancing manufacturability of hybrid organic–inorganic semiconductors for large area optoelectronics (AMYS). A.W. thanks Quentin Guesnay from the PV-LAB at EPFL for his initial help with the evaporation system. A.W. thanks Sasa Vranjkovic and the Empa tool shop for their help in the technical construction and manufacturing of CNC machined parts. A.W. further thanks Yousuf Hemani from the Laboratory of Advanced Analytical Methods at Empa for lending a Si photodiode used to calibrate the stress light intensity. A.W. also thanks Kerstin Thorwarth from the Laboratory of Surface Science & Coating Technologies for technical support on the optical characterization setup. Lastly, A.W. thanks Monalisa Ghosh from the Laboratory of Surface Science & Coating Technologies for technical support on the climate chamber.

# Supporting Information

Scheme of the optical pathway within the automated UV-Vis setup, spectrum of the white light used for degradation, temporal stability and spatial uniformity of the aging conditions, supplementary XRD diffractograms and optical spectra, further details about the machine-learning model, and training as well as testing data distribution.

*Supporting Information for*

# Advancing Combinatorial Aging Studies of Hybrid Perovskite Thin-Films *via* Precise Automated Characterization Methods and Machine Learning Augmented Analysis

Alexander Wieczorek[a], Austin G. Kuba[b], Jan Sommerhäuser[a], Luis Nicklaus Caceres[a], Christian Wolff[b,*], Sebastian Siol[a,*]

[a]*Laboratory for Surface Science and Coating Technologies,*
*Empa − Swiss Federal Laboratories for Materials Science and Technology, Switzerland.*

[b]*Institute of Electrical and Microengineering (IEM), Photovoltaic and Thin-Film Electronics Laboratory,*
*EPFL − École Polytechnique Fédérale de Lausanne, Switzerland.*

*Corresponding authors: christian.wolff@epfl.ch, sebastian.siol@empa.ch

# Table of Figures

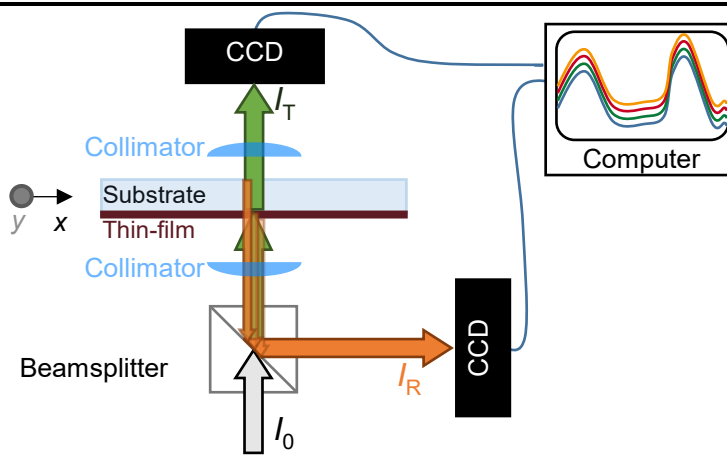

**Figure S 1.** Schematic of the beam path and position of the beam splitter and collimators. The light intensity of the light source $I_0$ is attenuated through the sample transmission or reflection.

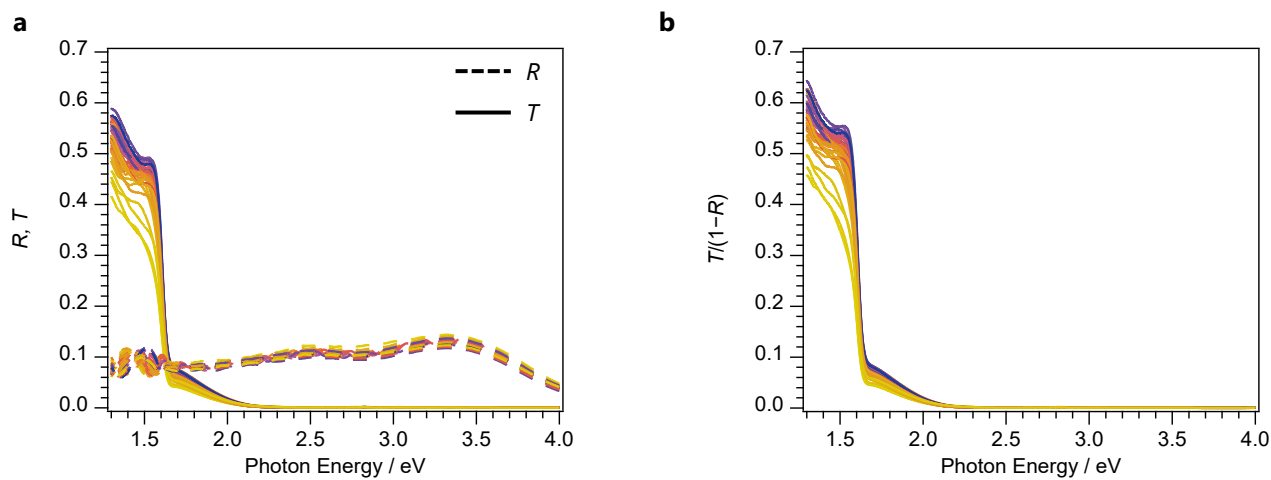

**Figure S 2.** a) Measured reflectance and transmittance of a combinatorial library, showing interference fringes around the bandgap of approx. 1.6 eV. b) Resulting reflection corrected transmittance.

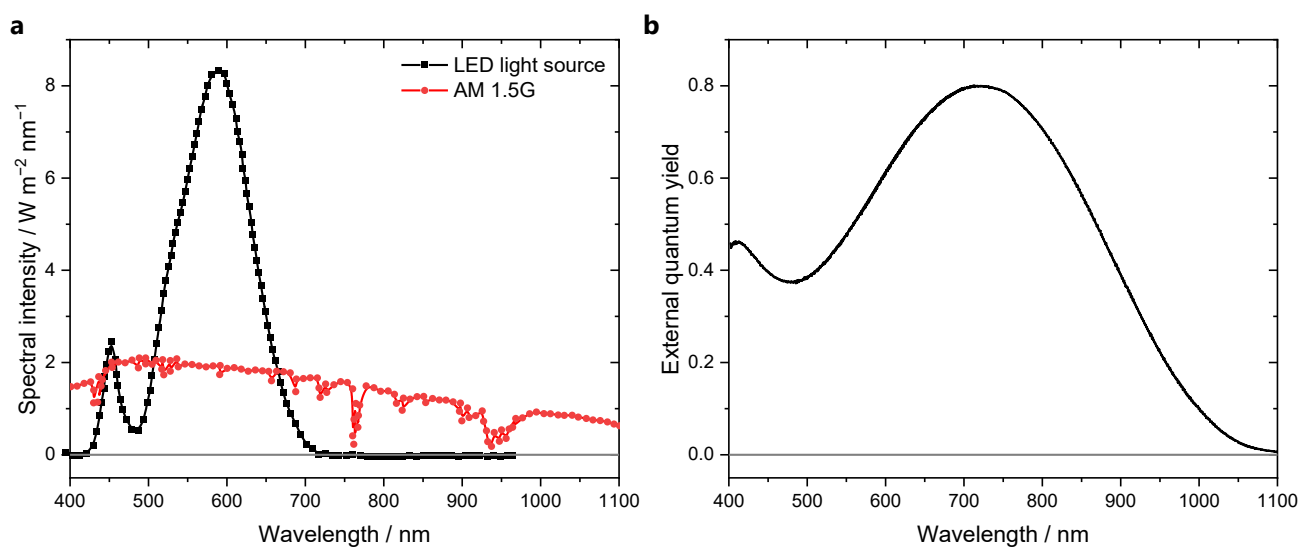

**Figure S 3.** A) Calibrated spectral intensity of the LED light source used against the AM 1.5G spectrum. B) Spectral external quantum yield (EQY) of the photodiode used to set the light intensity of the LED light source.

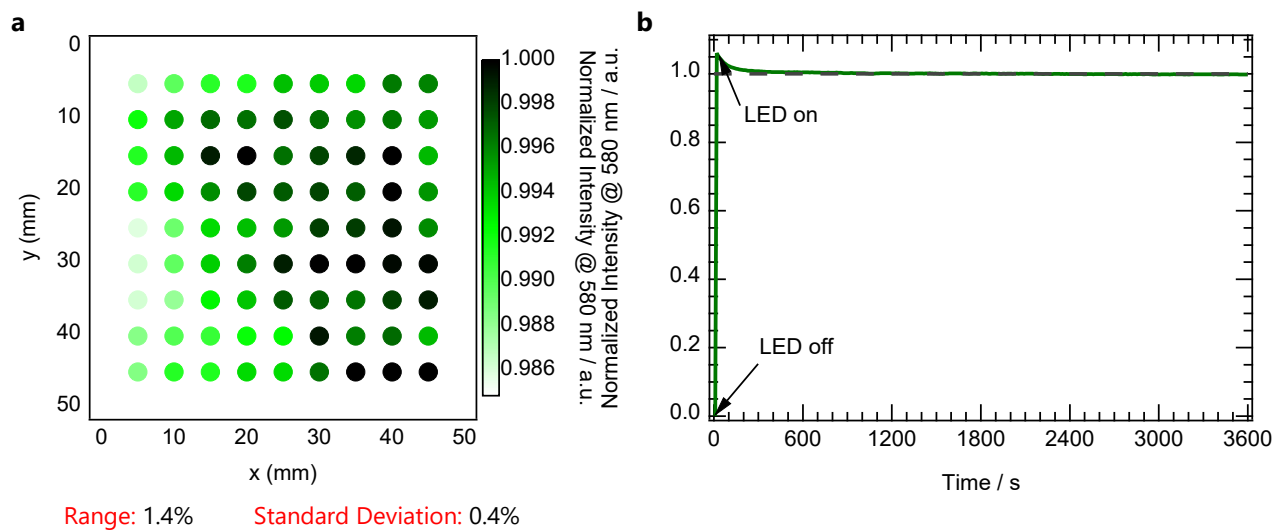

Range: 1.4%    Standard Deviation: 0.4%

**Figure S 4.** A.) Spatial homogeneity of the stress light across the library area. B.) Temporal stability of the stressing light. Following a slight intensity overshoot of up to 6%, an intensity plateau is reached after approx. 6 min which corresponds to the calibrated light intensity. The signal intensity at 580 nm was selected in both cases as the emission maximum of the used LED light source.

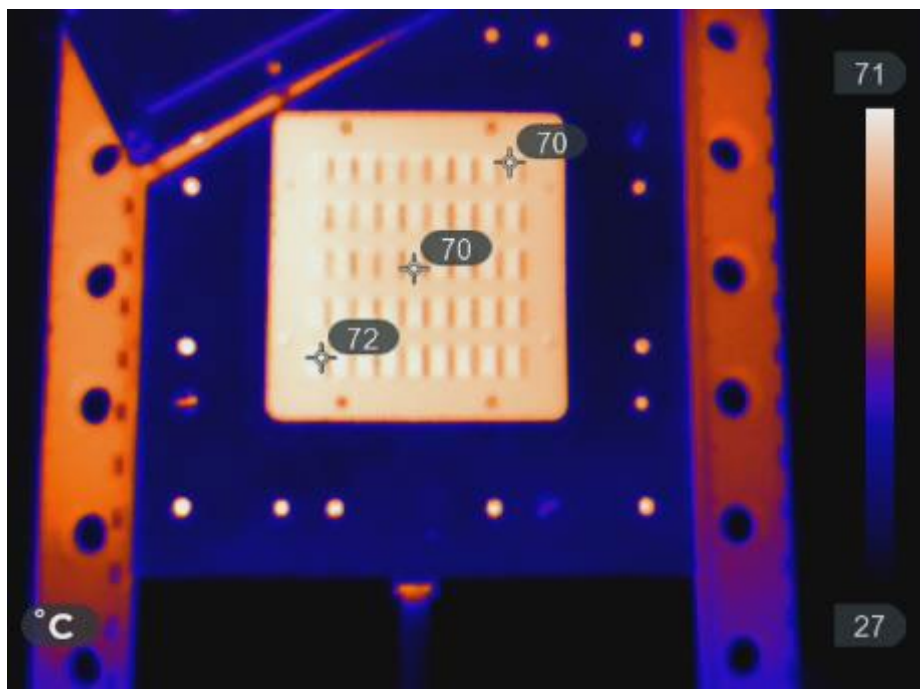

**Figure S 5.** False-colour thermal image of climate chamber with marked temperatures across the combinatorial grid, showing a standard deviation of <1 °C on perovskite thin-film library. An offset from the actual temperature of 85 °C resulted from the low emissivity of the samples on the substrate.

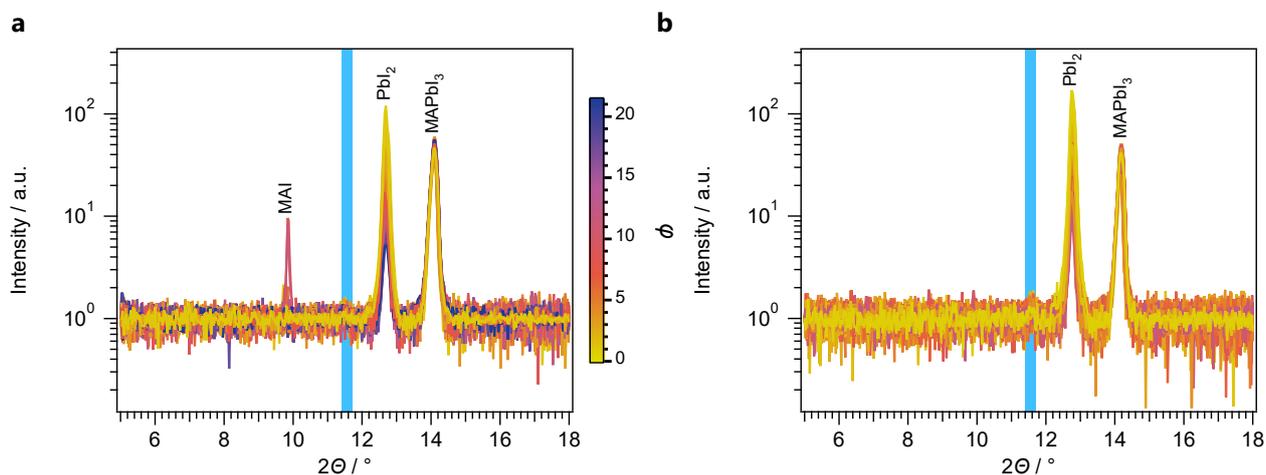

**Figure S 6.** 2Θ XRD measurements at shallower angles performed a) before and b) after the aging procedure. One sample exhibited a feature at approx. 9.8 °, indicative of residual unreacted MAI on the thin-film[1] which was however absent in the aged thin-films. Overconverted perovskite thin-films (*i.e.* those with overstochiometric amounts of A-site cations) can typically be detected at these shallower diffraction angles.[2,3] However, no similar feature could be detected at the angle previously reported for overconverted MAPbI$_3$ thin-films (blue marker).[1]

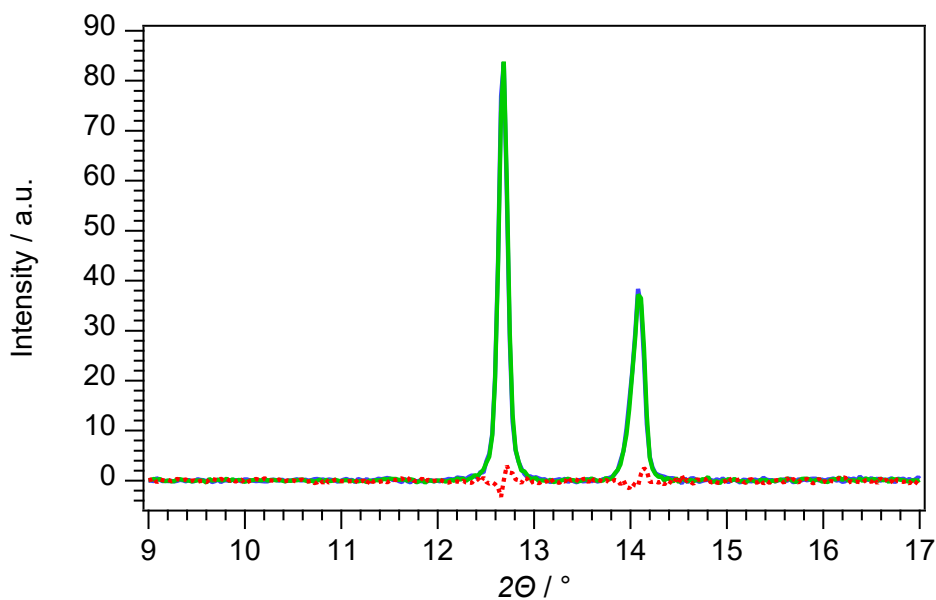

**Figure S 7.** XRD diffractogram of library samples measured within the inert-gas XRD Dome before (blue lines) and after (green lines) 17 h within the Dome. No significant change of the peaks was observed based on the differential spectrum (red dots).

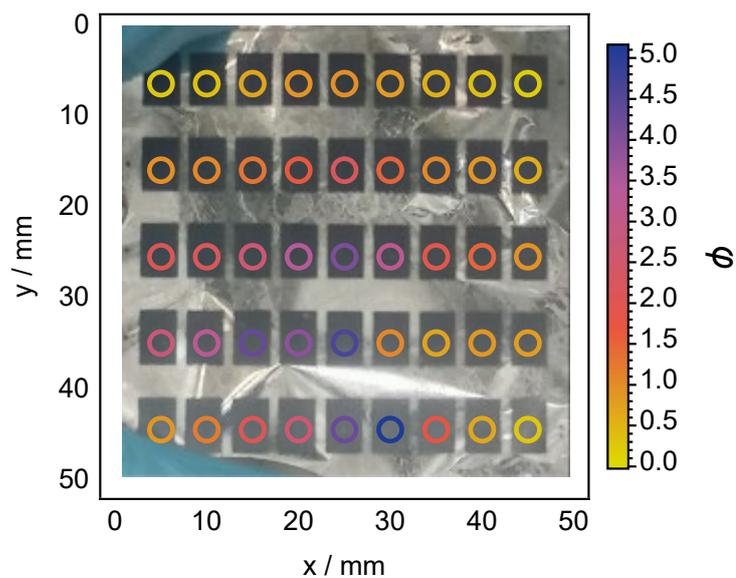

**Figure S 8.** Overlayed mapping of the MHP/PbI$_2$ phase ratio on top of a photograph of the pristine library. Behind the patterned library on the transparent substrate, an aluminum foil can be seen.

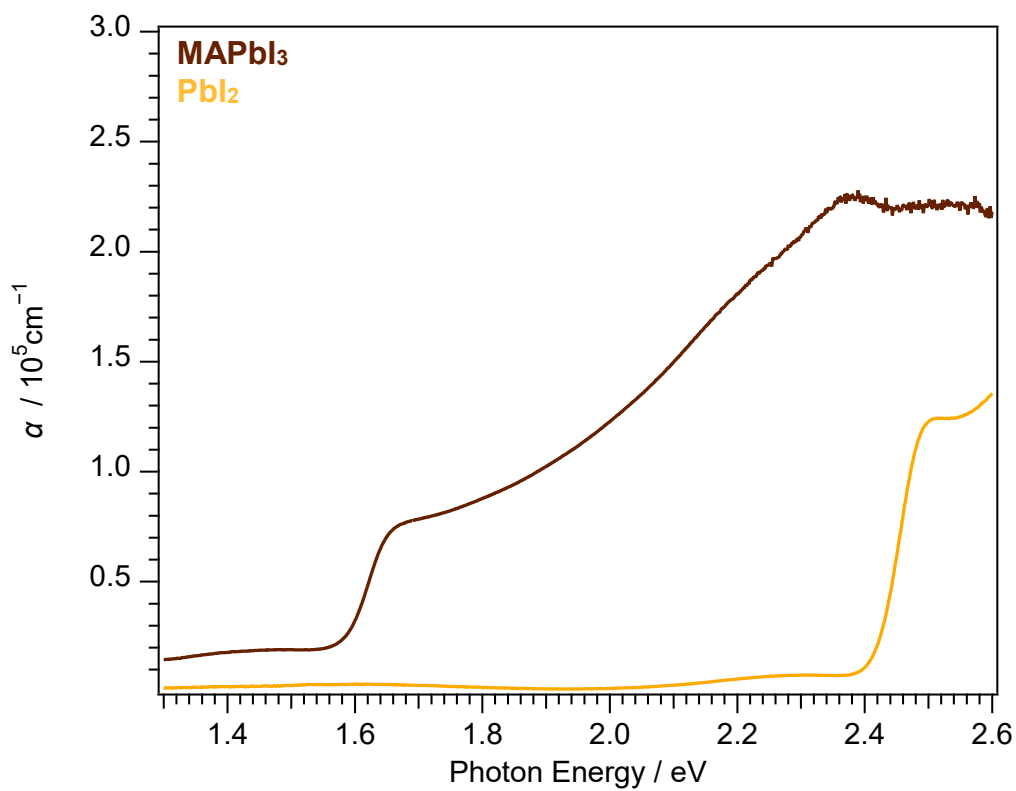

**Figure S 9.** Absorption coefficient *α* vs Photon Energy for a thermally evaporated PbI$_2$ template and a resulting MAPbI$_3$ thin-film with a high conversion rate with band gaps of approx. 2.4 eV and 1.6 eV, respectively.

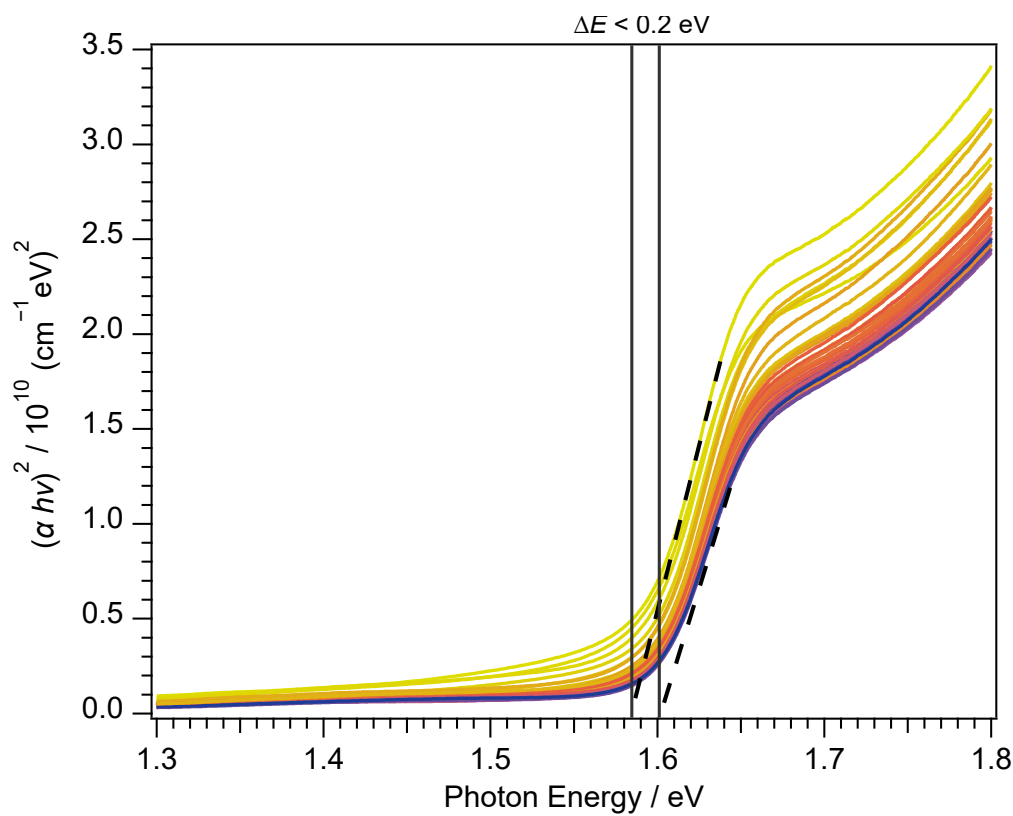

**Figure S 10.** Tauc plot analysis for an indirect allowed transition with varying *φ*. A shift of less than 0.2 eV can be observed in the extracted band gaps from samples with the highest and lowest *φ* values.

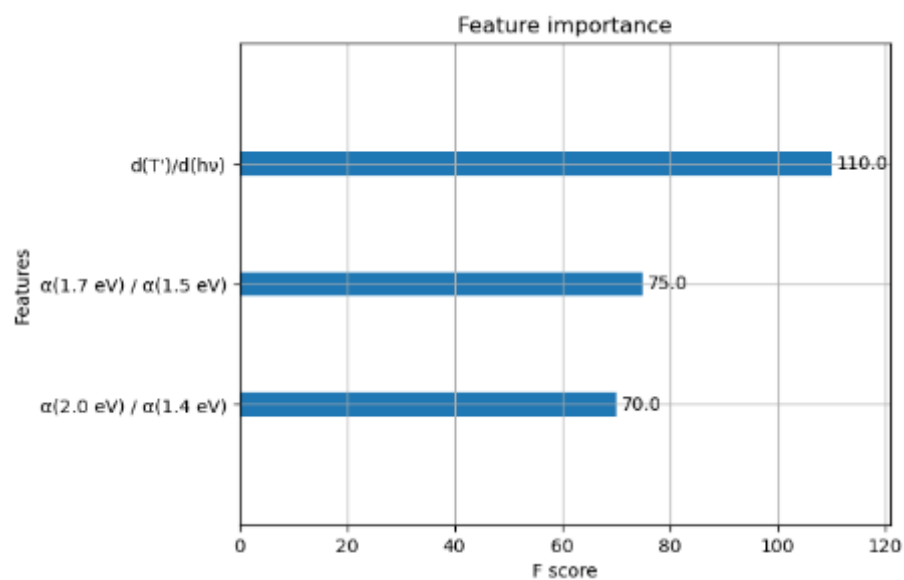

**Figure S 11.** Feature importance of the optimized XGBRegressor model.

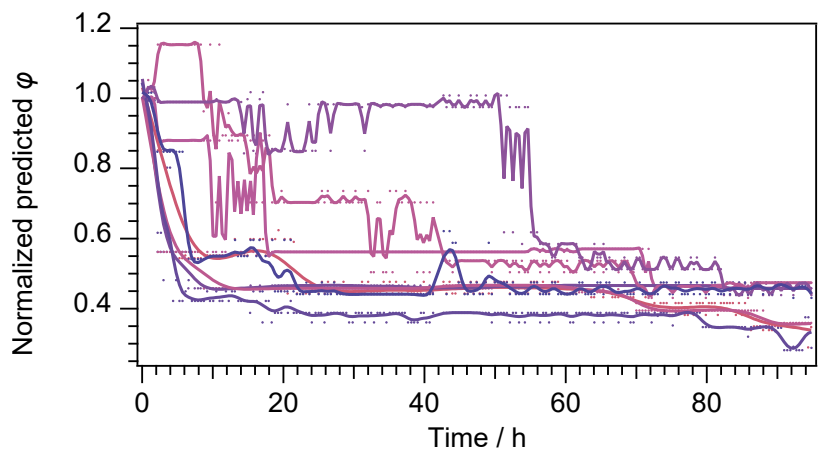

**Figure S 12.** Regressor model applied on samples with high *φ* values of > 2.5.

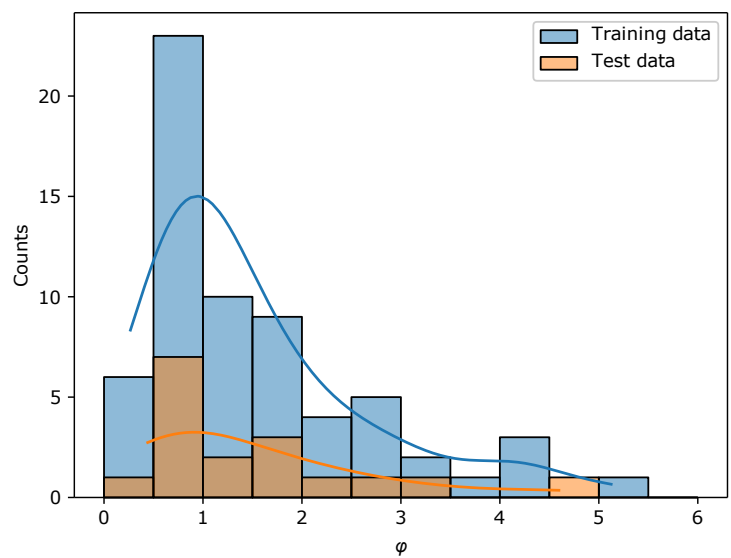

**Figure S 13.** Distribution of MHP/PbI$_2$ phase ratio within the training and test data expressed as a histogram with an overlayed probability distribution function. Samples with a MHP/PbI2 phase ratio of *φ* > 2.5 constitute the minority, amounting for only 20% of the total training data. The distribution function exhibites the similar shape for both data sets, indicating a likewise represention of the biased distribution within both the training and test data.